\documentclass[11pt]{article}

\usepackage{amssymb,amsmath,amsfonts,amsthm,lscape}

 \usepackage{cite}
 \usepackage{hyperref}
\hypersetup{
   colorlinks   =  true,
    linkcolor    = blue,
    citecolor    = red,
     urlcolor	=magenta,     
}

\newcommand\be{\begin{equation}}
\newcommand\ee{\end{equation}}
\newcommand\bea{\begin{eqnarray}}
\newcommand\eea{\end{eqnarray}}

 \def\mf {\mathfrak}

 \newcommand\dd{{\rm d}}

  \newcommand\st{{\rm st} }%spacetime
 \newcommand\til{{\rm tl}} %time-like

\DeclareMathOperator\spn{span}

\def\>#1{{\bf #1}}

\begin{document}

\thispagestyle{empty}

\

 \vskip1cm

\noindent {\Large{\bf {Noncommutative spacetimes  versus noncommutative spaces\\[6pt] of geodesics
}}}

\medskip
\medskip
\medskip
\medskip

 \begin{center}

 {\sc {Francisco  J  Herranz\footnote{
 Based on the contribution presented at ``The XII International Symposium on Quantum Theory and Symmetries" (QTS12),
July 24--28, 2023, 
 Czech Technical University in Prague, Czech Republic.\\ To appear in Journal of Physics: Conference Series.}, Angel Ballesteros$^{1}$, Giulia Gubitosi$^{2}$\\[4pt]  and Ivan Gutierrez-Sagredo$^{3}$}}

 \end{center}

\medskip

\noindent
{$^1$ Departamento de F\'isica, Universidad de Burgos, 
E-09001 Burgos, Spain}  \\[2pt]
\noindent
{$^2$ Dipartimento  di  Fisica  Ettore  Pancini,  Universit\`a  di  Napoli  Federico  II,  and  INFN,  Sezione  di  Napoli,  Complesso  Univ.~Monte S.~Angelo, I-80126 Napoli, Italy}\\[2pt]
{$^3$ Departamento de Matem\'aticas y Computaci\'on, Universidad de Burgos, 
E-09001 Burgos, Spain}
 
\medskip

\noindent
 {E-mails:\quad {\tt  fjherranz@ubu.es, angelb@ubu.es,  giulia.gubitosi@unina.it,\\[2pt]  igsagredo@ubu.es}}

\begin{abstract}  
\noindent
The aim of this contribution is twofold. First, we show that when two (or more) different quantum groups share the same noncommutative spacetime, such an `ambiguity' can be resolved by considering together their corresponding noncommutative spaces of geodesics. In any case, the latter play a mathematical/physical role by themselves and, in some cases, they can be interpreted as deformed phase spaces.
Second, we explicitly show  that noncommutative spacetimes can be reproduced from `extended' noncommutative spaces of geodesics which are those enlarged by the time translation generator. These general ideas are described in  detail for the $\kappa$-Poincar\'e and $\kappa$-Galilei algebras.
\end{abstract}

%%%%%%%%%%%%%%%%%%%%%%%%%%%%%%%%%%%%%%%%%%%%%%%

%

%%%%%%%%%%%%%%%%%%%%%%%%%%%%%%%%%%%%%%%%%%%%%%%

%%%%%%%%%%%%%%%%%%%%%%%%%%%%%%%%%%%%%%%%%%%%%%%

\section{Introduction}

It is well-known that  there exist quantum deformations of different Lie algebras that share a common noncommutative space with an underlying homogeneous space. This fact occurs for Riemannian, pseudo-Riemannian and semi-Riemannian spaces~\cite{GH2021symmetry}. Probably, the most paradigmatic situation concerns the $\kappa$-Minkowski~\cite{Maslanka1993},  $\kappa$-Galilei~\cite{GKMMK1992,BGGH2020kappanewtoncarroll}, and $\kappa$-Carroll~\cite{CK4d,BGGH2020kappanewtoncarroll} spacetimes, all of them   defined by the same  Lie algebraic structure. Our first goal is to show that this kind of `ambiguity' can be resolved when looking at their corresponding noncommutative space of geodesics, which clearly distinguishes among such a common noncommutative spacetime when both structures are considered simultaneously~\cite{BGGH2023}.

This issue is explained in detail by considering the 
 $\kappa$-Poincar\'e and $\kappa$-Galilei algebras. In particular,  in the next section we describe the (3+1)D homogeneous spacetimes and 6D spaces of geodesics for any kinematical algebra~\cite{BLL}. We then  apply these structures to the Poincar\'e and Galilei groups. The construction of the  corresponding noncommutive spacetimes and spaces of (time-like) geodesics are addressed for the  $\kappa$-Poincar\'e and  $\kappa$-Galilei  deformations in sections~\ref{s3} and \ref{s4}, respectively. We remark that for the  $\kappa$-Galilei case, its noncommutative space of geodesics leads to a  homogeneous quadratic algebra which can be interpreted as a deformed phase space in a natural way; in fact, this resembles  Snyder's model~\cite{Snyder1947}.

Our second aim is to show that it is possible to obtain a (3+1)D noncommutative spacetime, not from the 6D noncommutative space of geodesics but from its  `extended' version. This is a 7D  noncommutative space which includes the quantum coordinate dual to the translation generator of the spacetime~\cite{BGGH2023}. This point is fully described in the last section. In particular, we  explicitly  present the  nonlinear changes of quantum coordinates that allows us to obtain the $\kappa$-Minkowski and $\kappa$-Galilei spacetimes from the   seven  quantum  coordinates of their `extended' noncommutative space of geodesics.

 %%%%%%%%%%%%%%%%%%%%%%%%

\section{Kinematical homogeneous spacetimes and spaces of geodesics}
\label{s2}

To start with, we recall the construction of  (3+1)D symmetric homogeneous spacetimes and 6D spaces of geodesics with constant curvature. Let $\mf g$ be any 10D kinematical Lie algebra appearing in the classification obtained in~\cite{BLL} and let $G$ its   corresponding Lie group. Thus this covers, among others, (A)dS, Poincar\'e, Newton--Hooke, Galilei, etc.
Then we consider the usual kinematical basis
\be
\mf g=\spn\{P_0,P_a, K_a, J_a\},\qquad a=1,2,3,
\label{a2}
\ee
corresponding to the generators of time and  space translations, boosts and rotations. 
  As a vector space,  $\mathfrak{g}$ can be written as the   sum of two subspaces through a Cartan decomposition:
\be
{\mathfrak{g}}=  \mathfrak{t} \oplus \mathfrak{h} , \qquad  [\mathfrak{h} ,\mathfrak {h} ] \subset \mathfrak{h}  .
\label{a3}
\ee

A generic kinematical homogeneous space  can be constructed as a  left coset space $G/H$, where $H$ is the Lie group of   $\mf h$. The isotropy subgroup $H$ leaves invariant a   point on  $G/H$,  the origin, so that the generators spanning $\mathfrak{h}$ act as rotations around it. The generators belonging to $\mathfrak{t}$ move the origin behaving as translations on $G/H$.  The most relevant   spaces $G/H$ correspond to spacetimes and spaces of geodesics, namely
 \be
\begin{array}{lll}
\multicolumn{3}{l}{ \!\!\!\!\! \mbox{$\bullet$ The (3+1)D   spacetime} \ \mathcal{S}  =G/H_\st\!:}\\[3pt]
\mathfrak g= \mathfrak t_\st \oplus \mathfrak h_\st ,  \quad & \mathfrak{t}_\st = \spn \{P_0,  {P_a} \}    ,\quad  & \mathfrak h_\st = \spn\{ {K_a},  {J_a} \}  .\\[6pt]
\multicolumn{3}{l}{ \!\!\!\!\! \mbox{$\bullet$ The 6D space of (time-like) lines} \ \mathcal{W}  =G/H_\til\!:}\\[3pt]
\mathfrak g= \mathfrak t_\til \oplus \mathfrak h_\til ,  \quad & \mathfrak{t}_\til = \spn \{  {P_a},  {K_a} \}    ,\   & \mathfrak h_\til = \spn\{P_0,  {J_a} \}=\mathbb R\oplus \mathfrak{so}(3) . 
 \end{array}
\label{a4} 
\ee
In fact, these are also symmetric homogeneous spaces since they are associated with the parity $\mathcal {P}$ and the time-reversal $\mathcal{T}$   involutive automorphisms  defined by~\cite{BLL}
\be
\mathcal{P}( P_0,P_a, K_a, J_a)=  (P_0,-P_a,-K_a, J_a),\qquad 
 \mathcal{T}( P_0, P_a, K_a, J_a)=  (-P_0, P_a,-K_a, J_a) .
 \label{a5} 
 \ee
Spacetimes are related to the composition $\mathcal{P}\mathcal{T}$ while spaces of geodesics to $\mathcal{P}$. A detailed description of the two spaces (\ref{a4}) for all  kinematical and Riemannian groups can be found in~\cite{GH2021symmetry} (cf.~section 5); note also that it is possible to   construct other types of kinematical homogeneous spaces, such as 6D  spaces of space-likes lines, 6D spaces of 2-planes and 4D spaces of 3-hyperplanes~\cite{GH2021symmetry}.

Although in this work we restrict ourselves  to consider the spaces (\ref{a4}) for the Poincar\'e and Galilean cases, we remark  that these ideas can be applied in a similar way to other kinematical/Riemannian spaces. 

The commutation rules of the Poincar\'e algebra in the basis (\ref{a2}) with $c=1$ are given by
\be
\begin{array}{llll}
[J_a,J_b]=\epsilon_{abc}J_c ,& \quad [J_a,P_b]=\epsilon_{abc}P_c , &\quad
[J_a,K_b]=\epsilon_{abc}K_c ,  &\quad  [J_a,P_0]=0 , \\[4pt]
\displaystyle{
  [K_a,P_0]=P_a  } , &\quad\displaystyle{[K_a,P_b]=\delta_{ab} P_0} ,    &\quad\displaystyle{[K_a,K_b]=-\epsilon_{abc} J_c} , 
 &\quad 
[P_\mu,P_\nu]=0 ,
\end{array}
\label{a6}
\ee
where hereafter Latin indices run as $a,b=1,2,3$, while Greek ones as $\mu=0,1,2,3$ and
   sum over repeated indices is assumed.  The Galilei algebra can then be obtained from (\ref{a6}) as the non-relativistic limit $c\to\infty$~\cite{Inonu:1953sp} or speed-space contraction~\cite{BLL} associated to the parity automorphism $\mathcal{P}$ (\ref{a5}) by  introducing explicitly the speed of light parameter $c$  in    (\ref{a6})  in the form
\be
 P_a\to     c^{-1}  P_a,\qquad   K_a\to   c^{-1}  K_a.
\label{a7}
\ee
The limit $c\to \infty$  leads  to the commutation relations defining the Galilei algebra:
\be
\begin{array}{llll}
[J_a,J_b]=\epsilon_{abc}J_c ,& \quad [J_a,P_b]=\epsilon_{abc}P_c , &\quad
[J_a,K_b]=\epsilon_{abc}K_c ,  &\quad  [J_a,P_0]=0 , \\[4pt]
\displaystyle{
  [K_a,P_0]=P_a  } , &\quad\displaystyle{[K_a,P_b]=0} ,    &\quad\displaystyle{[K_a,K_b]= 0 } , 
 &\quad 
[P_\mu,P_\nu]=0 .
\end{array}
\label{a8}
\ee
  Recall that $H_\st$ is  just the Lorentz group ${\rm SO}(3,1)$ for the Poincar\'e case, while 
it  is isomorphic to the 3D Euclidean group ${\rm ISO}(3)$  for    the Galilei   group.

Then we introduce coordinates on the spaces (\ref{a4})  through their metric structures. For the flat spacetime $\mathcal{S}$ (either $\mathcal{M}$ or  $\mathcal{G}$) one can define  Cartesian coordinates $x^\mu$ dual to the generators $P_\mu$ yielding the well-known expressions:  
\be
\begin{array}{ll}
\mbox{$\bullet$ Minkowski $\mathcal{M}$:}& {\rm d} s^2=( \dd x^0)^2-( \dd x^1)^2-  ( \dd x^2)^2-( \dd x^3)^2   .\\[4pt]
\mbox{$\bullet$ Galilei $\mathcal{G}$:}& {\rm d} s_{(1)}^2=( \dd x^0)^2 ,\quad {\rm d} s_{(2)}^2= ( \dd x^1)^2+ ( \dd x^2)^2+( \dd x^3)^2 \ \ \mbox{on $x^0=$ const.}  
 \end{array}
\label{a9} 
\ee
In the Galilei spacetime the `main' metric $g^{(1)}$ is degenerate,   i.e.~the `absolute-time'  $x^0$,  which generates a foliation   whose leaves are defined at  a constant time. A `subsidiary'  3D   Euclidean spatial metric   $g^{(2)}$ is restricted to each leaf of the foliation. The quantum analogue of $x^\mu$ will be    the $\hat x^\mu$ generators shown below in their corresponding noncommutative spacetimes. 
 
 The metric structure on the space of (time-like) geodesics or worldlines $ \mathcal{W} $ (\ref{a4})  is less known;  see~\cite{HS1997phasespaces} for details. We denote by 
$(  y^a,  \eta^a)$ the six coordinates dual to $P_a$ and $K_a$  on this space. The $y^a$ can be interpreted as position-type coordinates and $\eta^a$  as momentum-type ones (or velocities). The metric is always degenerate and there arises an invariant foliation~\cite{HS1997phasespaces,BGH2019worldlinesplb}:
\be
\begin{array}{ll}
\mbox{$\bullet$ Poincar\'e $ \mathcal{W} $:}&  {\rm d} s_{(1)}^2=(\cosh\eta ^2)^2 (\cosh\eta^3)^2(\dd \eta^1)^2+ (\cosh\eta^3)^2 (\dd \eta^2)^2+( \dd \eta^3 )^2 ,\\[4pt]
 &  {\rm d} s_{(2)}^2=( \dd y^1)^2+  ( \dd y^2)^2+( \dd y^3)^2 ,\ \ \mbox{on $\eta^a=$ constant}  .\\[8pt]
\mbox{$\bullet$ Galilei  $ \mathcal{W} $:}&  {\rm d} s_{(1)}^2=  (\dd \eta^1)^2+  (\dd \eta^2)^2+( \dd \eta^3 )^2 ,\\[4pt]
 &  {\rm d} s_{(2)}^2=( \dd y^1)^2+  ( \dd y^2)^2+( \dd y^3)^2 ,\ \ \mbox{on $\eta^a=$ constant}  . 
   \end{array}
\label{a10} 
\ee
In the   Poincar\'e $ \mathcal{W} $, the `main' metric $g^{(1)}$ determines the 3-velocity space and it  is a  hyperbolic Riemannian metric of negative constant curvature   (i.e.~$-1/c^2$). The geodesic distance $\chi$ 
 \be
\cosh\chi=\cosh\eta ^1 \cosh\eta ^2 \cosh\eta^3\, 
 \label{a11}
 \ee
 corresponds to the relative rapidity between an observer at rest and one with a uniform motion with velocity $\eta^a$.   In the  Galilei $ \mathcal{W} $, the `main' metric $g^{(1)}$ reduces to the usual one for the space of velocities in Newtonian mechanics. The geodesic distance in  the 3-velocity space now reads  
 \be
\chi^2 \equiv \boldsymbol{\eta}^2 =  (  \eta^1)^2+  (  \eta^2)^2+(   \eta^3 )^2  .
 \label{a12}
 \ee
  
In the next two sections, we review the construction of the noncommutative counterparts of the two spaces (\ref{a4}) in terms of  quantum coordinates $\hat x^\mu$ and $(\hat  y^a, \hat \eta^a)$ for Poincar\'e and Galilei.

%%%%%%%%%%%%%%%%%%%%%%%%%%%%%%%%%%%%%%%%%%%%%%%

 \section{Noncommutative Minkowki spacetime and Poincar\'e  space of geodesics}
\label{s3}

We consider the usual $\kappa$-Poincar\'e algebra~\cite{LRNT1991,GKMMK1992,LNR1992fieldtheory,Maslanka1993,MR1994,Zakrzewski1994poincare} and from it we construct the corresponding 
 $\kappa$-Minkowki space~\cite{Maslanka1993} and the $\kappa$-Poincar\'e  space of time-like geodesics~\cite{BGH2019worldlinesplb}. We remark that the approach followed in~\cite{BGH2019worldlinesplb} can be applied to any other homogeneous space $G/H$ and quantum deformation under certain conditions; these require to start from a  coisotropic Poisson homogeneous space and a coboundary Lie bialgebra (see~\cite{GH2021symmetry} and references therein). In particular, such a procedure has recently been  used to obtain all  noncommutative spaces of $\kappa$-Poincar\'e geodesics in~\cite{BGH2022} as well as   all 
 (3+1)D noncommutative (A)dS and Minkowski  spacetimes that preserve a quantum Lorentz subgroup in~\cite{Ballesteros:2021inq} (these are quite different from $\kappa$-deformations).
     
     Explicitly, let
 $\rho : \mathfrak g  \rightarrow \text{End}(\mathbb R ^5)$ be a faithful matrix representation  for a generic element $X\in \mathfrak g $ of the   Poincar\'e   algebra  (\ref{a6})    given by 
\begin{equation}
\rho(X)=   x^\mu \rho(P_\mu)  +  \xi^a \rho(K_a) +  \theta^a \rho(J_a) =
\left(\begin{array}{ccccc}
0&0&0&0&0\cr 
x^0 &0&\xi^1&\xi^2&\xi^3\cr 
x^1 &\xi^1&0&-\theta^3&\theta^2\cr 
x^2 &\xi^2&\theta^3&0&-\theta^1\cr 
x^3 &\xi^3&-\theta^2&\theta^1&0
\end{array}\right)  .
\label{a13}
\end{equation}
The Poincar\'e group  is obtained by
    exponentiation. According to the Cartan decomposition (\ref{a4})  for the (3+1)D Minkowski spacetime  $ \mathcal{M} =G/H_\st $, the  element of the group is chosen as
\be
\begin{array}{l}
 G_\mathcal{M}= \exp{\!\bigl(x^0 \rho(P_0)\bigr)} \exp{\!\bigl(x^1 \rho(P_1)\bigr)} \exp{\!\bigl(x^2 \rho(P_2)\bigr)} \exp{\!\bigl(x^3 \rho(P_3)\bigr)} \, H_\st,   \\[5pt]
 H_\st= \exp\bigl({\xi^1\! \rho(K_1)}\bigr) \exp\bigl({\xi^2\! \rho(K_2)}\bigr) \exp\bigl({\xi^3 \!\rho(K_3)}\bigr) \exp\bigl({\theta^1 \!\rho(J_1)}\bigr) \exp\bigl({\theta^2 \!\rho(J_2)} \bigr)\exp\bigl({\theta^3\! \rho(J_3)}\bigr) .
 \end{array}
 \label{a14}
\ee
In this way $x^\mu$ provide a well-defined set of coordinates in $ \mathcal{M} $.
From $ G_\mathcal{M}$ we compute  left- and right-invariant vector fields and   introduce them into  the  Sklyanin bracket~\cite{ChariPressley1994} determined by the $\kappa$-Poincar\'e classical $r$-matrix~\cite{Maslanka1993,Zakrzewski1994poincare}:
\be
r= {\kappa}^{-1} (K_1 \wedge P_1 + K_2 \wedge P_2 + K_3 \wedge P_3) .
\label{a15}
\ee
The    Sklyanin brackets for the classical coordinates $x^\mu$ determine the Poisson version of   $\kappa$-Minkowski  and turn out to be given by linear brackets that can be  quantized directly, thus providing the well-known $\kappa$-Minkowski    $ \mathcal{M}_\kappa$ in terms of quantum coordinates $\hat x^\mu$~\cite{Maslanka1993}:
 \be
[\hat x^a,\hat x^0]= {\kappa}^{-1}\, \hat x^a,\qquad [\hat x^a,\hat x^b]=0 .
\label{a16}
\ee

 For the 6D Poincar\'e space of time-like geodesics  $\mathcal{W}  =G/H_\til$ (\ref{a4}), 
we can consider again the same matrix representation (\ref{a13}) but  with the following ordering of the exponentials:
\be
\begin{array}{l}
G_{\mathcal{W} } = \exp\bigl({\eta^1 \rho(K_1)}\bigr) \exp\bigl({y^1 \rho(P_1)}\bigr) \exp\bigl({\eta^2 \rho(K_2)}\bigr) \exp\bigl({y^2 \rho(P_2)} \bigr) \\[5pt]
\qquad\qquad\quad  \times  \exp\bigl({\eta^3 \rho(K_3)}\bigr) \exp\bigl({y^3 \rho(P_3)}\bigr) \,H_{\til} ,   \\[6pt]
H_{\til} = \exp\bigl( {\phi^1 \rho(J_1)}\bigr) \exp\bigl( {\phi^2 \rho(J_2)}\bigr) \exp\bigl( {\phi^3 \rho(J_3)}\bigr) \exp\bigl( {y^0 \rho(P_0)} \bigr).
 \end{array}
 \label{a17}
\ee
Therefore $( y^a, \eta^a)$ constitute an appropriate set of coordinates  in $\mathcal{W}$. 
Then from $G_{\mathcal{W} }$ we   derive    left- and right-invariant vector fields  and obtain the Poisson brackets for $( y^a, \eta^a)$ via the  Sklyanin bracket with  (\ref{a15}). 
It turns out that, although they are no longer linear, they can be  quantized straightforwardly since there are no order ambiguities.  The resulting   $\kappa$-Poincar\'e    space  of time-like geodesics $\mathcal{W}_\kappa$ is defined in terms of the quantum coordinates $( \hat y^a, \hat \eta^a)$ by~\cite{BGH2019worldlinesplb}:
\begin{equation}
\begin{split}
\big[\hat y^1, \hat y^2\big] &= {\kappa}^{-1} \left(  \sinh \hat \eta^1\, \hat y^2 -\frac{ \tanh \hat \eta^2}{\cosh \hat \eta^3}\, \hat y^1\right) ,\\[2pt]
\big[\hat y^1,\hat y^3\big] &=  {\kappa}^{-1} \big( \sinh  \hat \eta^1\, \hat y^3  - \tanh  \hat \eta^3\,\hat y^1 \big), \\[2pt]
\big[\hat y^2, \hat y^3\big] &= {\kappa}^{-1}\big( \cosh  \hat \eta^1  \sinh  \hat \eta^2   \, \hat y^3 - \tanh  \hat \eta^3\,\hat y^2\big), \\[2pt]
\big[\hat y^1, \hat \eta^1\big] &= {\kappa}^{-1}\frac{ \bigl(\cosh  \hat \eta^1  \cosh  \hat \eta^2  \cosh  \hat \eta^3 -1\bigr)}{\cosh  \hat \eta^2  \cosh  \hat \eta^3 }, \\[2pt]
\big[\hat y^2, \hat \eta^2\big] &= {\kappa}^{-1} \frac{ \bigl(\cosh  \hat \eta^1  \cosh  \hat \eta^2  \cosh  \hat \eta^3 -1\bigr)}{\cosh  \hat \eta^3 }, \\[2pt]
\big[\hat y^3, \hat \eta^3\big] &=  {\kappa}^{-1}\bigl(\cosh  \hat \eta^1  \cosh  \hat \eta^2  \cosh  \hat \eta^3 -1\bigr) \, , \\[2pt]
\big[\hat \eta^a, \hat\eta^b\big] &= 0,\quad  \forall a,b,\qquad \big[\hat y^a,\hat \eta^b\big] = 0,\quad    a \neq b. 
\end{split}
\label{a18}
\end{equation}
Observe that  the three momentum-type quantum coordinates $\hat \eta^a$ commute with each other.  Consequently, it is quite natural to   search  for a differential realization
of  $(\hat y^a,\hat \eta^a)$, fulfilling (\ref{a18}),   as operators acting  on the space of functions $\Psi(\eta^1,\eta^2,\eta^3)$ (hence  on   an underlying 3-velocity space with classical coordinates $\eta^a$). The resulting  differential realization  is found be~\cite{BGGH2023}:
\begin{align}
\begin{split}
&\hat y^1  \Psi  = \frac{1}{\kappa} \,\frac{ \bigl( \cosh  \chi -1\bigr)}{\cosh    \eta^2  \cosh    \eta^3 }\,   \frac{\partial \Psi}{\partial \eta^1}  ,\qquad 
 \hat y^2  \Psi  = \frac{1}{\kappa}\, \frac{ \bigl(\cosh  \chi -1\bigr)}{\cosh    \eta^3 } \,\frac{\partial \Psi}{\partial \eta^2}  ,\\[3pt]
&\hat y^3  \Psi  =  \frac{1}{\kappa}\, \bigl(\cosh  \chi-1\bigr) \frac{\partial \Psi}{\partial \eta^3}   ,\qquad 
  \hat \eta^a \Psi = \eta^a\Psi ,
\end{split}
\label{a19}
\end{align}
where $\cosh  \chi$ is given in   (\ref{a11}). In addition, this result allows us to describe $\mathcal{W}_\kappa$   in terms of   `quantum Darboux operators' $(\hat q^a,\hat p^a)$ defined by~\cite{BGH2019worldlinesplb,BGGH2023}
\be
\hat q^1     := \frac{\cosh  \hat  \eta^2  \cosh   \hat \eta^3 }{ \cosh\hat  \chi -1 }\,   \hat y^1  ,\quad\ \  
 \hat q^2  :=  \frac{\cosh    \hat\eta^3 }{  \cosh \hat \chi -1 } \,\hat y^2 ,\quad \ \ 
\hat q^3     :=  \frac 1 {\cosh \hat \chi-1 }\, \hat y^3   ,\quad\  \ 
  \hat p^a :=\hat  \eta^a  ,
\label{a20}
\ee
with $\cosh\hat \chi=\cosh\hat\eta ^1 \cosh\hat\eta ^2 \cosh\hat\eta^3$ (see (\ref{a11})). This means that  $(\hat q^a,\hat p^a)$  satisfy  the canonical commutation relations
\begin{align}
\big[ \hat q^a,\hat q^b\big]= \big[\hat p^a,\hat p^b\big]= 0, \qquad   \big[ \hat q^a,\hat p^b\big]= {\kappa}^{-1} \, \delta_{ab}  .
\label{a21}
\end{align}
Therefore,   $   {\mathcal{W} }_\kappa$ can be expressed as   three copies of the   Heisenberg--Weyl  algebra   of quantum mechanics where the deformation parameter $\kappa^{-1}$ replaces the Planck constant $\hbar$.   We also recall that  a first phenomenological analysis for  $   {\mathcal{W} }_\kappa$ (\ref{a18})  has been    performed in~\cite{BGGM2021fuzzy}, which may have relevance in a quantum gravity setting~\cite{Addazi:2021xuf}.

 %%%%%%%%%%%%%%%%%%%%%%%%%%%%%%%%%%%%%%%%%%%%%%%

  \section{Noncommutative Galilei spacetime and Galilei  space of\\ geodesics}
\label{s4}

As far as   $\kappa$-Galilei algebra is concerned,  recall that it is a non-coboundary deformation, that is, its Hopf structure has no   underlying classical $r$-matrix~\cite{BGHOS1995quasiorthogonal}. Certainly, one can consider the $r$-matrix (\ref{a15}), since it is valid for the Galilei algebra (\ref{a8}), but this would lead to a trivial cocommutator~\cite{BGGH2020kappanewtoncarroll}, so there is no deformation and the Hopf structure remains trivial. The proper $\kappa$-Galilei algebra was formerly obtained in~\cite{GKMMK1992} as the non-relativistic limit of the $\kappa$-Poincar\'e algebra without keeping a classical $r$-matrix (see also~\cite{Azcarraga95}). Therefore, the approach of the previous section is precluded.

In terms of the Lie bialgebra contraction approach~\cite{BGHOS1995quasiorthogonal}, there exists a fundamental non-coboundary contraction from the $\kappa$-Poincar\'e algebra  to 
the $\kappa$-Galilei algebra which, obviously, makes use of the usual  Lie algebra transformation (\ref{a7}) but requires to preserve the quantum deformation parameter $\kappa$ unchanged. 
By taking into account such a contraction, the $r$-matrix (\ref{a15})  diverges under   the limit $c\to \infty$, however  the Hopf structure has  a well-defined non-trivial limit providing the  $\kappa$-Galilei algebra~\cite{GKMMK1992,BGGH2020kappanewtoncarroll}. This contraction has a quantum group counterpart that enables one  to obtain the 
 $\kappa$-Galilei spacetime  $ \mathcal{G}_\kappa$ and $\kappa$-Galilei    space  of  geodesics  from the $\kappa$-Minkowski spacetime  $ \mathcal{M}_\kappa$ (\ref{a16})  and the $\kappa$-Poincar\'e    space  of time-like geodesics $\mathcal{W}_\kappa$ (\ref{a18}). By duality,  the contraction (\ref{a7}) implies that the quantum coordinates $\hat x^\mu$ and $(\hat y^a ,\hat \eta^a)$ must be transformed as~\cite{BGGH2023}
\be
\hat x^a\to c\, \hat x^a,\qquad \hat y^a\to c\, \hat y^a,\qquad \hat \eta^a\to c\, \hat \eta^a,
\label{a22}
\ee
preserving $\hat x^0$ and $\kappa$.

 Then, if we apply the map (\ref{a22}) to  $ \mathcal{M}_\kappa$ (\ref{a16}) and take the limit $c\to\infty$ we  obtain  exactly the same expressions for $ \mathcal{G}_\kappa$ as for $ \mathcal{M}_\kappa$. Nevertheless, 
under this contraction the cumbersome expressions (\ref{a18}) defining the noncommutative Poincar\'e space of geodesics reduce to the following homogeneous quadratic algebra: 
\begin{equation}
\begin{split}
\big[\hat y^a, \hat y^b\big] &= {\kappa}^{-1} \big(  \hat \eta^a\hat y^b   - { \hat \eta^b\hat y^a  } \big) , \qquad \big[\hat \eta^a, \hat\eta^b\big] = 0,\quad  \forall a,b ,\\[2pt]
\big[\hat y^a, \hat \eta^b\big] &= \tfrac12  {\kappa}^{-1} \, \delta_{ab}  \left(  ( \hat \eta^1)^2+  ( \hat \eta^2)^2+(  \hat \eta^3 )^2 \right)  ,
\end{split}
\label{a23}
\end{equation}
which defines the noncommutative Galilei space of geodesics. Consequently, at the quantum group level, the $\kappa$-Poincar\'e and $\kappa$-Galilei deformations can be clearly distinguished from each other if both noncommutative spacetimes and  spaces of geodesics are taken into account together.
 
Furthermore, the quadratic expressions (\ref{a23}) can be rewritten in terms of the operators
\be
\hat L_a=\epsilon_{abc}  \, { \hat y^b \hat \eta^c }\, , \qquad  \hat { \boldsymbol{\eta}}^2=
  ( \hat \eta^1)^2+  ( \hat \eta^2)^2+(  \hat \eta^3 )^2    ,
\label{a24}
\ee
yielding
\be
\big[\hat y^a, \hat y^b\big] = -{\kappa}^{-1} \epsilon_{abc}\,\hat L_c,\qquad
\big[\hat y^a, \hat \eta^b\big] =   \tfrac12  {\kappa}^{-1} \, \delta_{ab}\, \hat { \boldsymbol{\eta}}^2,
\qquad  \big[\hat \eta^a, \hat\eta^b\big] = 0.
\label{a25}
\ee
Provided that   $(\hat y^a,\hat \eta^a)$ can be considered as  quantum position- and momentum-type  variables,  $\hat L_a$ can be interpreted as a  quantum angular momenta-type operator and $  \hat { \boldsymbol{\eta}}^2$ as the quantum kinetic energy operator; the latter is the quantum analogue of (\ref{a12}). Therefore, under such an interpretation, the noncommutative Galilei space of geodesics can alternatively be regarded as a deformed quantum  phase space. 
  In this respect, we remark that  the algebra (\ref{a25}) closely resembles  Snyder's model~\cite{Snyder1947} in such a manner that the former is   algebraically similar to a   Snyder--Euclidean phase space~\cite{BGH2020snyder,BGH2020snyderPOS},  also known in the literature as a  `non-relativistic' Snyder model~\cite{Mignemi:2011gr, Lu:2011it, Mignemi:2012gr,Ivetic:2015cwa}. Note that the first commutator in (\ref{a25}) would correspond to a Snyder--Euclidean space with the curvature determined by the quantum deformation parameter through  the factor $( -{\kappa}^{-1} )$.

 The differential realization of  $(\hat y^a,\hat \eta^a)$ on the space of functions $\Psi(\eta^1,\eta^2,\eta^3)$, now satisfying (\ref{a23}), can be obtained either from the Poincar\'e one (\ref{a19}) applying the contraction (\ref{a22}), or by direct computation; this reads as
 \begin{align}
\begin{split}
& \hat y^a  \Psi  = \frac{1}{2\kappa} \,{ \boldsymbol{\eta}}^2\,   \frac{\partial \Psi}{\partial \eta^a}  ,\qquad  \hat \eta^a \Psi = \eta^a\Psi ,
\end{split}
\label{a26}
\end{align}
 where $\boldsymbol{\eta}^2$ is given by (\ref{a12}).   And the corresponding quantum Darboux operators  $(\hat q^a,\hat p^a)$, fulfilling  (\ref{a21}) are given by
 \be
\hat q^a     := \frac{2 }{  \hat { \boldsymbol{\eta}}^2 }\,   \hat y^a  ,  \qquad  \hat p^a :=\hat  \eta^a  .
\label{a27}
\ee

%%%%%%%%%%%%%%%%%%%%%%%%%%%%%%%%%%%%%%%%%%%%%%%

\section{Noncommutative spacetimes from noncommutative spaces of geodesics?}
\label{s5}

So far we have shown that for different deformations sharing a common quantum spacetime, differences could arise when noncommutative  spaces of geodesics are also considered together.  Moreover, for some cases, noncommutative  spaces of geodesics could naturally be  interpreted as deformed phase spaces. Although we have restricted ourselves to deal with the $\kappa$-Poincar\'e and  $\kappa$-Galilei deformations, we remark that this idea is quite general. For instance,  the   $\kappa$-Carroll algebra~\cite{CK4d,BGGH2020kappanewtoncarroll}  also shares as    noncommutative spacetime   the same structure as the $\kappa$-Minkowski one (\ref{a16}). But in~\cite{BGGH2023} it has also been shown how the noncommutative Carroll space of lines distinguishes itself from both the Poincar\'e and Galilei cases.

Now a natural question arises:  is there any relationship between the  noncommutative spacetime and the space of geodesics coming from the same given quantum deformation? Clearly,    the answer is negative since from the construction of the associated homogeneous spaces (\ref{a4}) the translation subspace  $ \mathfrak{t}_\til = \spn \{  {P_a},  {K_a} \}$ does not include $ \mathfrak{t}_\st = \spn \{P_0,  {P_a} \}  $.  However, this problem can be circumvented by enlarging the 6D space of geodesics $\mathcal{W} $ up to include the time generator $P_0$. The resulting  7D  homogeneous space $G/H$ is called the  `extended' space of geodesics $ \widetilde {\mathcal{W} } =G/\widetilde{H}_\til$ which is provided by the following Cartan decomposition~\cite{BGGH2023}
\be
\mathfrak g= \widetilde{\mathfrak t}_\til \oplus \widetilde{\mathfrak h}_\til ,  \qquad  \widetilde{\mathfrak{t}}_\til = \spn \{ P_0, {P_a},  {K_a} \}    ,\qquad   \widetilde{\mathfrak h}_\til = \spn\{  {J_a} \}= \mathfrak{so}(3) .
\label{a28} 
\ee
This, in turn, means that the classical coordinates for $ \widetilde {\mathcal{W} } $ include $y^0$ dual to $P_0$ besides the former  $(  y^a,  \eta^a)$ dual to $(P_a,K_a)$.
One could therefore expect to recover the (3+1)D noncommutative spacetime from
the noncommutative extended space of geodesics for a given quantum group.

 This idea was developed in detail for the particular $\kappa$-Poincar\'e and  $\kappa$-Galilei deformations in~\cite{BGGH2023}. In what follows we summarize the main results.
 
 The  $\kappa$-Poincar\'e  extended  space  of  geodesics $ \widetilde {\mathcal{W} }_\kappa $  is defined by the commutators (\ref{a18}) along with six additional  commutators $[\hat y^a,\hat y^0]$ and $[\hat \eta^a,\hat y^0]$. Following the procedure of section~\ref{s3}, we compute the remaining Poisson brackets for the classical coordinates,  finding that there are ordering ambiguities. Nevertheless, their quantization can finally be  performed by setting the ordering $(\hat \eta^a)^m\,(\hat y^a)^n$, namely
   \begin{equation}
\begin{split}
&\big[\hat y^1, \hat y^0 \big] = {\kappa}^{-1} \left( \hat y^1 -  \frac{\sinh \hat \eta^1 \tanh \hat \eta^2}{\cosh \hat \eta^3}\, \hat y^2 - \sinh \hat \eta^1 \tanh \hat \eta^3 \, \hat y^3 \right) ,\\[2pt]
&\big[\hat y^2, \hat y^0 \big] =  {\kappa}^{-1} \left( \hat y^2 +\frac{\sinh \hat \eta^1 \tanh \hat \eta^2}{\cosh \hat \eta^3}\,  \hat y^1 - \cosh \hat \eta^1 \sinh \hat \eta^2 \tanh \hat \eta^3 \, \hat y^3  \right) ,\\[2pt]
&\big[\hat y^3, \hat y^0 \big] = {\kappa}^{-1}\left( \hat y^3 +  \sinh \hat \eta^1 \tanh \hat \eta^3\,\hat y^1 + \cosh \hat \eta^1 \sinh \hat \eta^2 \tanh \hat \eta^3 \,  \hat y^2\right) ,\\[2pt]
&\big[\hat \eta^1, \hat y^0 \big] =  {\kappa}^{-1}\, \frac{\sinh \hat \eta^1}{\cosh \hat \eta^2 \cosh \hat \eta^3} ,\\[2pt] 
&\big[\hat \eta^2, \hat y^0 \big] =  {\kappa}^{-1}\, \frac{\cosh \hat \eta^1 \sinh \hat \eta^2}{\cosh \hat \eta^3} ,\\[2pt]
&\big[\hat \eta^3, \hat y^0 \big] =  {\kappa}^{-1} \cosh \hat \eta^1 \cosh \hat \eta^2 \sinh \eta^3  .
\end{split}
\label{a29}
\end{equation}
The  differential realization (\ref{a19}) is completed with
\be
 \hat y^0  \Psi  = -\frac 1\kappa \left( \frac{\sinh    \eta^1}{\cosh    \eta^2  \cosh    \eta^3 }\,  \frac{\partial \Psi}{\partial \eta^1} + \frac{\cosh    \eta^1 \sinh    \eta^2 }{ \cosh    \eta^3 }\,\frac{\partial \Psi}{\partial \eta^2} +   \cosh    \eta^1 \cosh    \eta^2  \sinh    \eta^3   \,  \frac{\partial \Psi}{\partial \eta^3}   \right).
\label{a30}
\ee

Thus we have all the structures to obtain   the $\kappa$-Minkowski spacetime  (\ref{a16}) 
from $ \widetilde {\mathcal{W} }_\kappa $  given by (\ref{a18}) and (\ref{a29}). This is achieved by means of the following nonlinear change of quantum coordinates   $ \hat x^\mu =( \hat y^\mu, \hat\eta^a)$:
\begin{align}
\begin{split}
&\hat x^0 = \cosh \hat \eta^1  \cosh \hat \eta^2  \cosh \hat \eta^3\, \hat y^0 + \sinh \hat \eta^1 \, \hat y^1 + \cosh \hat \eta^1  \sinh \hat \eta^2\, \hat y^2 + \cosh \hat \eta^1  \cosh \hat \eta^2  \sinh \hat \eta^3\, \hat y^3  ,\\[2pt]
&\hat x^1  =\sinh \hat \eta^1 \cosh \hat \eta^2  \cosh \hat \eta^3\, \hat y^0 +  \cosh \hat \eta^1 \, \hat y^1+   \sinh \hat \eta^1  \sinh \hat \eta^2 \, \hat y^2+ \sinh \hat \eta^1 \cosh \hat \eta^2  \sinh \hat \eta^3\, \hat y^3   ,\\[2pt]
&\hat  x^2  = \sinh \hat \eta^2  \cosh \hat \eta^3\, \hat y^0 +  \cosh \hat \eta^2\,\hat y^2  +   \sinh \hat \eta^2  \sinh \hat \eta^3\, \hat y^3, \\[2pt]
&\hat x^3  =  \sinh \hat \eta^3 \, \hat y^0+ \cosh \hat \eta^3\,  \hat y^3 \, .
\end{split}
\label{a31}
\end{align}
 
Likewise the  $\kappa$-Galilei  extended  space  of  geodesics in section~\ref{s4} is defined by the previous commutators (\ref{a23}) together with 
 \begin{equation}
 \big[\hat y^a, \hat y^0 \big] =  {\kappa}^{-1} \, \hat y^a ,\qquad   \big[\hat \eta^a, \hat y^0 \big] =  {\kappa}^{-1}\,  \hat \eta^a  .
\label{a32}
\end{equation}
And the differential realization   (\ref{a26}) requires to add
\be
\hat y^0  \Psi  = -\frac 1\kappa \left(     \eta^1 \,  \frac{\partial \Psi}{\partial \eta^1} +     \eta^2  \,\frac{\partial \Psi}{\partial \eta^2} +    \eta^3   \,  \frac{\partial \Psi}{\partial \eta^3}   \right).
\label{a33}
\ee
The $\kappa$-Galilei spacetime  $\mathcal{G}_\kappa$ (\ref{a16})  is recovered  from the quantum extended space of geodesics via  the change of quantum variables
 \begin{equation}
 \hat x^0 = \hat y^0 ,\qquad   \hat x^a = \hat y^a  +  \hat \eta^a      \hat y^0   .
\label{a34}
\end{equation}
 
 Observe that as a byproduct of the Poincar\'e differential realization (\ref{a19}) and (\ref{a30}), one can  obtain directly a differential realization of the  quantum coordinates $\hat x^\mu$, expressed as in (\ref{a31}),  on the same 3D classical momentum space $\eta^a$. And similarly for the Galilean deformation. More details can be found in~\cite{BGGH2023}.

To conclude, it is worth remarking that  the approach and ideas described in this contribution can be applied to other types of $\kappa$-Poincar\'e deformations~\cite{BP2014extendedkappa}, thus also covering the so called `space-like' and `light-like' (or null-plane) deformations. In fact,  all    $\kappa$-Poincar\'e spaces of geodesics (for the three types) have been recently constructed in~\cite{BGH2022} but without their `extended' version. Furthermore, to the best of our knowledge, a construction of   noncommutative (A)dS spaces of geodesics  has not  yet been performed. Work along these lines is currently in progress.

%%%%%%%%%%%%%%%%%%%%%%%%%%%%%%%%%%%%%%%%%%%%%%%

\section*{Acknowledgments}

This work has been partially supported by Agencia Estatal de Investigaci\'on (Spain)  under grant  PID2019-106802GB-I00/AEI/10.13039/501100011033, by the Regional Government of Castilla y Le\'on (Junta de Castilla y Le\'on, Spain) and by the Spanish Ministry of Science and Innovation MICIN and the European Union NextGenerationEU  (PRTR C17.I1). The authors would like to acknowledge the contribution of the European Cooperation in Science and Technology COST Action CA18108.

%%%%%%%%%%%%%%%%%%%%%%%%%%%%%%%%%%%%%%%%%%%%%%%

%\section*{References}

\end{document}